# Preventive and Reactive Cyber Defense Dynamics Is Globally Stable

Ren Zheng, Wenlian Lu, and Shouhuai Xu

*Abstract*—The recently proposed *cybersecurity dynamics* approach aims to understand cybersecurity from a holistic perspective by modeling the evolution of the global cybersecurity state. These models describe the interactions between the various kinds of cyber defenses and the various kinds of cyber attacks. We study a particular kind of cybersecurity dynamics caused by the interactions between preventive and reactive defenses (e.g., filtering and malware detection) against push- and pull-based cyber attacks (e.g., malware spreading and "drive-by download" attacks). The dynamics was previously shown to be globally stable in a *special* regime of the parameter universe, but little is known beyond this special regime. In this paper, we resolve an open problem in this domain by proving that the dynamics is globally stable in the *entire* parameter universe (i.e., the dynamics always converges to a unique equilibrium). We discuss the cybersecurity meanings and implications of this theoretic result. We also prove that the dynamics converges *exponentially* to the equilibrium except for a special parameter regime, in which case the dynamics converges *polynomially*. Since it is often difficult to compute the equilibrium, we propose new bounds of the equilibrium and numerically show that these bounds are tighter than those proposed in the literature.

*Index Terms*—Cybersecurity dynamics, preventive and reactive cyber defense dynamics, cyber attack-defense dynamics, cybersecurity models, cybersecurity foundation

## I. Introduction

Any approach that aims to understand cybersecurity from a holistic perspective needs to model the interactions between the various kinds of cyber attacks and the various kinds of cyber defenses. Recently, a new approach called *cybersecurity dynamics* [30] has shown its potential for systematically understanding, characterizing, and quantifying cybersecurity from a holistic perspective [39], [33], [28], [31], [8], [5], [35], [18], [36], [29], [34], [15]. At a high level, this approach aims to characterize how the attack-defense interactions govern the evolution of the global cybersecurity state [30], [31], and how the resulting characteristics can be applied to guide cyber defense operations (see, for example, [18], [35]). The approach was inspired by multiple earlier endeavors in several disciplines [30], including: (i) *Biological Epidemic Models* [19], [14], [2], [1], [9] as well as their adaptations to the cyberspace setting, namely *Cyber Epidemic Models*, which were pioneered by Kephart and White [12], [13] and later elegantly developed to accommodate specific kinds of network structures (e.g., power-law [22], [20], [22], [23], [21], [3]) and *arbitrary* network structures (e.g., [27], [7], [4], [26]); (ii) Interacting Particle Systems [17], which study the collective behaviors and phenomena that can be exhibited by interacting components; (iii) Microfoundation in Economics [11], which aims to connect macroeconomic theories to the underlying microeconomic behaviors of agents. The cybersecurity dynamics approach has two telling features that distinguish it from these inspiring endeavors. One feature is that the approach can offer a systematic treatment of complex cyber attack-defense interactions, including preventive and reactive defense dynamics [34], [5], [36], [29], [15], adaptive defense dynamics [35], active defense dynamics [39], [33], [18], and proactive defense dynamics [8]. Another feature is that the approach articulates a systematic set of technical barriers that need to be adequately addressed [30].

In this paper, we investigate *preventive and reactive defense dynamics*, namely the evolution of the global cybersecurity state caused by the interaction between cyber attacks and preventive and reactive cyber defenses, which represent two classes of cyber defense mechanisms that have been widely employed. Preventive defense mechanisms, including various intrusion prevention tools such as filtering, aim to prevent cyber attacks from succeeding. Reactive defense mechanisms, such as anti-malware tools, aim to detect and clean the compromised computers. The cyber attacks we consider include the following two classes. One class is the push-based attacks, such as malwares that actively seek and attack vulnerable computers in cyberspace. The other class is the pull-based attacks, such as drive-by downloads [24] by which vulnerable browsers/computers get compromised when visiting malicious websites. The full-fledged preventive and reactive defense dynamics model accommodating these attack-defense interactions was introduced in [16] and analytically treated in [34]. The full-fledged model supersedes the model that was investigated in [27], [7], [4], [26], which considered push-based attacks only.

An important research problem for understanding (preventive and reactive) cyber defense dynamics is: What phenomenon does the dynamics exhibit? The state-of-the-art understanding is that the dynamics is globally stable in a *special* regime of the parameter universe, but little is known beyond this special regime [27], [7], [4], [26], [34]. Figure 1(a) illustrates the special parameter regime as the white-colored *open* area, where "open" means that properties of the dynamics on the dashed boundary were not known. The illustration equally applies to the model of preventive and reactive cyber defenses against push-based attacks [27], [7], [4], [26], and the full-fledged model of preventive and reactive cyber defenses

Ren Zheng is with School of Mathematics, Fudan University, China and Department of Computer Science, University of Texas at San Antonio, USA. Wenlian Lu is with School of Mathematics, Fudan University, China. Shouhuai Xu is with Department of Computer Science, University of Texas at San Antonio, USA. (Correspondence: shxu@cs.utsa.edu)

against push-based and pull-based attacks [16], [34].

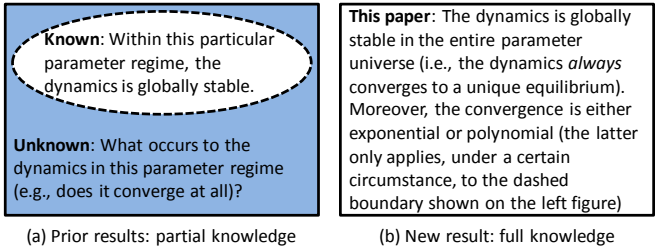

Fig. 1. Illustration of results. (a) Prior results: The dynamics is globally stable in the parameter regime that is illustrated as the white-colored *open* area of parameter regime. It was not known whether or not the dynamics converges in the other regime of the parameter universe, including the dashed boundary. (b) Our new result: The dynamics is globally stable in the *entire* parameter universe, and the convergence speed is completely characterized.

**Our contributions.** In this paper, we make three contributions. First, we tackle the open problem illustrated in Figure 1(a), by proving that preventive and reactive cyber defense dynamics is globally stable in the entire parameter universe as illustrated in Figure 1(b). This means that the dynamics *always* converges to a unique equilibrium regardless of the initial global cybersecurity state, as illustrated in Figure 2(a). In order to demonstrate the cybersecurity meaning of this theoretical result, let us consider the following security metric: the fraction $i$ of compromised computers in an enterprise network, namely the number of compromised computers divided by the total number of computers in the network. This metric reflects the global cybersecurity state, and can be used to set (for example) the threshold of successful attacks that can be tolerated (e.g., using threshold cryptosystems to tolerate the fraction of compromised computers [6], [32]). Note that this metric is not static but dependent upon time, namely that $0 \leq i(t) \leq 1$ for any $t \in [0, +\infty)$. Knowing that the dynamics is globally stable allows the defender to measure $i(t)$ when the dynamics is in equilibrium, possibly through the use of some sampling methods such as the one presented in [34]. This measurement-based method for estimating the equilibrium is valuable because it does not require the defender to know the values of the model parameters. In contrast, Figure 2(b) illustrates an "unmanageable" situation, where $i(t)$ does not exhibit any pattern and can even be chaotic, meaning that $i(t)$ is unpredictable because it is too sensitive to the initial value $i(0)$. Consequently, $i(t)$ cannot be measured in real time regardless of the sampling methods that may be used, because the state might have already changed after the sampling operation.

Second, it was known that the dynamics converges *exponentially* to a unique equilibrium in the white-colored open area of parameter regime illustrated in Figure 1(a) [27], [7], [4], [26], [34]. Little is known beyond this special parameter regime. We prove that the dynamics of preventive and reactive defenses against push-based attacks (i.e., $\alpha = 0$ in the terminology of the model) converges *polynomially* in the parameter regime corresponding to the dashed boundary illustrated in Figure 1(a). We also prove that the dynamics converges *exponentially* in the parameter regime corresponding to the blue-colored area illustrated in Figure 1(a). That is, we now have a full characterization of the convergence speed of the dynamics.

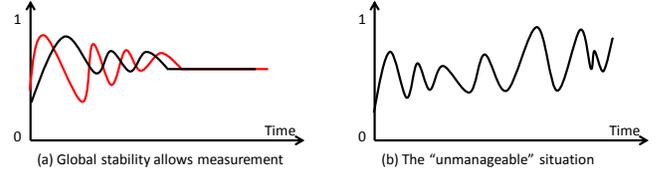

Fig. 2. Global stability implies that sampling-based security measurement is possible. (a) Global stability means that the dynamics always converges to a unique equilibrium that is independent of the initial cybersecurity state or the fraction $i(0) \in [0, 1]$ of compromised computers at time $t = 0$. In this case, the equilibrium $i^* = i(t \to \infty)$ can be measured using some sampling-based methods *without* knowing the values of the model parameters. (b) The "unmanageable" situation where $i(t) \in [0, 1]$ cannot be feasibly measured because the global cybersecurity state might have already changed after the measurement operation.

Third, we give new bounds on the equilibrium $i^* = i(t \to \infty)$. The bounds are useful because it is often difficult to compute the equilibrium. For example, when the defender cannot use the sampling-based approach to estimate the equilibrium, the defender can instead use the upper bound of $i^*$ for decision-making, while noting that the price is the possible overprovision of defense resources. We numerically show that our new upper and lower bounds are respectively tighter than their counterparts in [34].

The rest of the paper is organized as follows. In Section II, we describe the continuous-time preventive and reactive cyber defense dynamics model, which is in parallel to the discrete-time dynamics model investigated in [16], [34]. In Section III, we review some results that will be used later in the present paper. In Section IV, we present our new results. In Section V, we use simulation to validate the new results (due to the lack of real data). In Section VI, we discuss the related prior work. In Section VII, we conclude the paper with some open problems for future research. Due to space limitation, we defer proofs of most theorems to the Appendix.

## II. MODELING PREVENTIVE AND REACTIVE CYBER DEFENSE DYNAMICS

As in the discrete-time model [34], we consider the interactions between push-based attacks and pull-based attacks against preventive defenses (e.g., intrusion prevention) and reactive defenses (e.g., anti-malware tools). Push-based attacks naturally formulate a *cyber attack structure* that can be modeled as a network $G = (V, E)$, where $V = \{1, 2, \ldots, n\}$ is the set of computers in the network, $(u, v) \in E$ means that a compromised computer or node $u$ can launch push-based attacks directly against a secure but vulnerable computer or node $v$, and $G$ can be directed or undirected. (It is worth mentioning that the abstraction can be equally applied to finer granularities; e.g., $u$ represents a software component in a computer.)

Note that $G$ is *not* necessarily the underlying physical or communication network structure, except perhaps in scenarios such as sensor networks or email networks. In principle, $G$ can be extracted from network security configurations. For example, some computers or IP addresses are prohibited from communicating with some other computers or IP addresses. This kind of access restriction is widely employed in the physical world for protecting sensitive facilities (e.g., only authorized users can have access to a military base), and is important for alleviating a cyber attack-defense *asymmetry* described in [39], [33]. The extraction of $G$ requires having access to the data describing enterprise networks and their security configurations, and is therefore an orthogonal research problem that needs to be investigated separately. Given that the kind of data is hard to obtain for academic researchers, characterization studies—including the present paper—should not make any restrictions on the structure of $G$. In other words, we should accommodate *arbitrary* network structures for $G$.

The adjacency matrix of cyber attack structure $G$ is denoted by $A = [a_{vu}]_{n \times n}$ where $a_{vu} = 1$ if and only if $(u, v) \in E$. Because we focus on attacks launched by compromised computers against others, we naturally let $a_{vv} = 0$, which means that *privilege escalation* is not explicitly accommodated. Instead, a computer is always treated as compromised after it is penetrated and before it is cleaned up. Denote by $\deg(v)$ the (in-)degree of node $v$ in $G$, because $G$ can be directed or undirected. Note that $\deg(v) = |N_v|$ where $N_v = \{u \in V : (u, v) \in E\}$.

In parallel to the discrete-time model investigated in [34], we consider a continuous-time model. At any point in time $t$, a node $v \in V$ is in one of two states: *secure* (i.e., secure but vulnerable, denoted by "0") or *compromised* (denoted by "1"). Let $s_v(t)$ and $i_v(t)$ respectively denote the probability that node $v$ is *secure* and *compromised* at time $t$, where $s_v(t) + i_v(t) = 1$ for any $v \in V$ and any $t \geq 0$. The global cybersecurity state probability vector is $s(t) = [s_1(t), \cdots, s_n(t)]$, or equivalently $i(t) = [i_1(t), \cdots, i_n(t)]$.

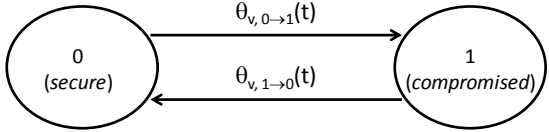

Fig. 3. State-transition diagram of an *individual* node $v \in V$ in the nonlinear Dynamical System model, where $\theta_{v,0 \to 1}$ is nonlinear.

Figure 3 illustrates the state-transition diagram of the continuous-time nonlinear Dynamical System model. Parameter $\theta_{v,1 \to 0}(t)$ describes the reactive defense power (i.e., effectiveness or capability), and is represented by $\beta \in (0, 1]$, namely the probability that a *compromised* node becomes *secure* at any point in time $t$ because of the reactive defense. The probability that a *secure* node gets *compromised*, namely $\theta_{v,0 \to 1}(t)$, is dependent upon the power of pull- and push-based cyber attacks against the preventive defense in question. For modeling the power of pull-based cyber attacks against the preventive defense, we use parameter $\alpha \in [0, 1]$ to denote the probability that a *secure* node becomes *compromised* at any time $t$ despite the deployed preventive defense. For modeling the power of push-based cyber attacks against the preventive defense, we use parameter $\gamma \in (0, 1]$ to denote the probability that a *compromised* node $u$ successfully attacks a *secure* node $v$, where $(u, v) \in E$. Assuming the *compromised* nodes launch attacks against their neighbors independent of each other, we have

$$\theta_{v,0 \to 1} = 1 - (1 - \alpha) \prod_{u \in N_v} \big(1 - \gamma i_u(t)\big).$$

This leads to the following continuous-time Dynamical System model for all $v \in V$:

$$\begin{cases} \frac{d}{dt} s_v(t) = \theta_{v,1 \to 0}(t) \cdot i_v(t) - \theta_{v,0 \to 1}(t) \cdot s_v(t) \\ \frac{d}{dt} i_v(t) = \theta_{v,0 \to 1}(t) \cdot s_v(t) - \theta_{v,1 \to 0}(t) \cdot i_v(t), \end{cases}$$

which can be rewritten as

$$\frac{d i_v(t)}{dt} = -\beta i_v(t) + \left[1 - (1-\alpha) \prod_{u \in N_v} \big(1 - \gamma i_u(t)\big)\right]\big(1 - i_v(t)\big). \tag{1}$$

The main research task is to analyze the system of $n$ nonlinear equations given by (1).

**Discussion on the assumptions and extensions**. First, we assume away the extraction of the cyber attack structure $G$ of real-world networks and the obtaining of the model parameters (i.e., $\alpha$, $\beta$ and $\gamma$). This is reasonable because these research problems are orthogonal to the focus of the present characterization study. Nevertheless, we note that characterization studies can suggest practical methods for alleviating the reliance on model parameters. For example, the characterization reported in [34] showed that it is possible to use sampling-based methods to estimate the equilibrium *without* knowing the values of $\alpha$, $\beta$ and $\gamma$. We anticipate that more results will be developed along this direction in the future.

Second, we assume $G$, $\alpha$, $\beta$, and $\gamma$ are time-independent. These assumptions would be valid for a short period of time, namely when the dynamics converges *exponentially*. It is an important future work to investigate the dynamics while accommodating time-dependent parameters, namely $G(t)$, $\alpha(t)$, $\beta(t)$ and $\gamma(t)$. On the other hand, we will show that our results can be easily extended to the setting where $\alpha$, $\beta$, and $\gamma$ are heterogeneous as follows: $\alpha_v$ accommodates that different nodes may have different capabilities in preventing pull-based attacks from succeeding, $\beta_v$ accommodates that different nodes may have different capabilities in detecting and cleaning up compromised computers, $\gamma_{uv}$ accommodates that different nodes may have different capabilities in preventing the push-based attacks launched from different neighbors from succeeding. In other words, the global stability result remains valid for heterogeneous $\alpha_v$, $\beta_v$ and $\gamma_{uv}$.

Third, we assume that the *compromised* nodes launch attacks independent of each other. This assumption is widely made in the literature. The first work that aimed at getting rid of this independence assumption (i.e., accommodating the dependence instead) is [28], which opens the door to many open problems for future study.

Despite the assumptions mentioned above, the present study is already fairly involved. Therefore, the present study should be seen as a significant step going beyond [34] in the direction towards systematically understanding preventive and reactive cyber defense dynamics. We believe that the present paper will inspire many future studies on, for example, weakening or eliminating the assumptions.

**Summary of notations**. Let $\mathbb{R}$ be the set of real numbers, $\mathbb{R}_+$ be the set of nonnegative real numbers, and $\mathbb{C}$ be the set of complex numbers. For comparing two $n$-dimensional vectors $i = [i_1, \ldots, i_n] \in \mathbb{R}^n$ and $j = [j_1, \ldots, j_n] \in \mathbb{R}^n$, we use the following notations:

- "$i \geq j$" means $i_v \geq j_v$ for all $v \in [1, \ldots, n]$,
- "$i > j$" means $i \geq j$ and $i_v > j_v$ for some $v \in [1, \ldots, n]$, and
- "$i \gg j$" means that $i_v > j_v$ for all $v \in [1, \ldots, n]$.

Table I summarizes the other major notations used in the paper.

TABLE I
NOTATIONS USED THROUGHOUT THE PAPER.

| | |
|---|---|
| $G = (V, E), A$ | the cyber attack structure $G$ and its adjacency matrix representation $A = [a_{vu}]_{n \times n}$ where $a_{vu} = 1$ if and only if $(u, v) \in E$ |
| $\alpha \in [0, 1]$ | the probability a *secure* but vulnerable node becomes *compromised* because of pull-based attacks despite the preventive defense |
| $\beta \in (0, 1]$ | the probability a *compromised* node becomes *secure* because of the reactive defense |
| $\gamma \in (0, 1]$ | the probability a *secure* but vulnerable node becomes *compromised* because of push-based attacks despite the preventive defense |
| $N_v$ | $v$'s neighbors; $N_v = \{u \in V : (u, v) \in E\}$ |
| $\deg(v)$ | the (in-)degree of node $v$; $\deg(v) = |N_v|$ |
| $\lambda_{M,1}$ | the largest eigenvalue of matrix $M$ |
| $i_v(t)$ | the probability $v$ is *compromised* at time $t$ |
| $\mathbb{0}$ | the original point $\mathbb{0} = [0, \cdots, 0]_n \in \mathbb{R}^n$ |
| $\mathbb{1}$ | the point $\mathbb{1} = [1, \cdots, 1]_n \in \mathbb{R}^n$ |

## III. PRELIMINARIES

### A. Algebraic Graph Theory

A directed graph, such as a directed cyber attack structure $G = (V, E)$, is said to be *strongly connected* if there is a directed path from any node to any other node. A *strongly connected component* is a strongly connected subgraph that is not contained in any other strongly connected subgraph. The adjacent matrix $A$ of graph $G$ can be rearranged to the Frobenius form

$$A = \begin{bmatrix} A_{11} & A_{12} & \cdots & A_{1p} \\ 0 & A_{22} & \cdots & A_{2p} \\ \vdots & \vdots & & \vdots \\ 0 & 0 & \cdots & A_{pp} \end{bmatrix} \quad (2)$$

where the strongly connected component $A_{jj}$ is irreducible for $j = 1, \ldots, p-1$, and $A_{pp}$ is irreducible or is a zero matrix with dimension 1. For $j = 1, \ldots, p$, we can define the subset $V_j \subseteq V$ corresponding to the strongly connected component $A_{jj}$ as

$$V_j = \{v \in V : v \text{ is the node corresponding to a row of } A_{jj}\}.$$

A strongly connected component $A_{jj}$ is said to be *without in-edges* if $A_{jk} = 0$ for all $k > j$.

### B. Cooperative Dynamical Systems

Our analysis will utilize some nice properties of cooperative dynamical systems and subhomogeneous maps.

*Definition 1: (cooperative dynamical system [10])*
Let $x = [x_1, \ldots, x_n]$. Let $f(\cdot) = [f_1(\cdot), \cdots, f_n(\cdot)]$ be a function vector in domain $\mathscr{D} \subset \mathbb{R}^n$. An autonomous dynamical system

$$\frac{\mathrm{d}x}{\mathrm{d}t} = f(x)$$

is said to be *cooperative* if $\partial f_v(x)/\partial x_u \geq 0$ holds for all $u, v \in \{1, \ldots, n\}$ and $u \neq v$.

*Definition 2: (subhomogeneity [38])* Let $f(\cdot) : \mathscr{D} \to \mathscr{D}$ be a continuous map in domain $\mathscr{D} \subset \mathbb{R}^n$.

- $f(\cdot)$ is said to be *subhomogeneous* if $f(\delta x) \geq \delta f(x)$ for any $x \in \mathscr{D}$ and $\delta \in (0, 1)$;
- $f(\cdot)$ is said to be *strictly subhomogeneous* if $f(\delta x) > \delta f(x)$ for any $x \in \mathscr{D}$ with $x \gg \mathbb{0}$ and $\delta \in (0, 1)$;
- $f(\cdot)$ is said to be *strongly subhomogeneous* if $f(\delta x) \gg \delta f(x)$ for any $x \in \mathscr{D}$ with $x \gg \mathbb{0}$ and $\delta \in (0, 1)$.

*Theorem 1: (part of Theorem 1.7 in [10])* If $f(\cdot)$ is cooperative in $\mathscr{D} \subset \mathbb{R}^n$, then $f(\cdot)$ is monotone in $\mathscr{D} \subset \mathbb{R}^n$, meaning that $f(j) \geq f(i)$ if $j > i$.

*Theorem 2: (Theorem 2.3.2 in [38])* Let $g(\cdot) : \mathscr{D} \to \mathscr{D}$ be a continuously differentiable map and $\mathscr{D} \subset \mathbb{R}^n_+$ be closed and contain $\mathbb{0}$. Suppose

(a) $g(\cdot)$ is monotone in $\mathscr{D}$, and
(b) $g(\cdot)$ is strongly subhomogeneous in $\mathscr{D}$.

If $g(\cdot)$ admits a nonempty compact invariant set $\mathcal{K} \subset \mathscr{D}$, then $g(\cdot)$ has a fixed point $x^* \gg \mathbb{0}$ such that every nonempty compact invariant set of $g(\cdot)$ in $\mathscr{D}$ consists of $x^*$.

*Theorem 3: (Theorem 2.2.6 in [38])* Let $f(\cdot) : \mathscr{D} \to \mathscr{D}$ be a continuously monotone map on $\mathscr{D}$. Assume that a nonempty compact invariant set $\mathcal{K} \subset \mathscr{D}$ contains only one equilibrium $x^*$. Then every trajectory attracted to $\mathcal{K}$ converges to $x^*$.

*Theorem 4: (Corollary 3.2 in [37])* Let $f(\cdot) : \mathscr{D} \to \mathscr{D}$ be a continuously differentiable map and $\mathscr{D} \subset \mathbb{R}^n_+$ be closed and contain $\mathbb{0}$. Suppose

(a) $f(\cdot)$ is cooperative in $\mathscr{D}$ and $\mathcal{D}f(x) = [\partial f_v(x)/\partial x_u]_{u,v \in V}$ is block-wise irreducible for every $x \in \mathscr{D}$,
(b) $f(\cdot)$ is strictly subhomogeneous in $\mathscr{D}$, and
(c) $f(\mathbb{0}) = 0$ and every positive trajectory of $f(\cdot)$ in $\mathscr{D}$ is bounded.

Let $s(\mathcal{D}f(\mathbb{0})) = \max\{\Re(\lambda) : \det(\lambda I - \mathcal{D}f(\mathbb{0})) = 0\}$.

- If $s(\mathcal{D}f(\mathbb{0})) \leq 0$, then $x^* = \mathbb{0}$ is globally asymptotically stable.
- If $s(\mathcal{D}f(\mathbb{0})) > 0$, there exists a unique equilibrium $x^* \in \mathscr{D} \setminus \{\mathbb{0}\}$ that is globally asymptotically stable.

### C. Previous Results

In the special case $\alpha = 0$ (i.e., preventive and reactive defenses against push-based attacks only), it is known that the dynamics converges to equilibrium $\mathbb{0}$ in the special regime of the parameter universe given by the following Theorem 5.

*Theorem 5: ([4])* Let parameters $G$, $\beta$ and $\gamma$ be specified as in the model described above. If

$$\lambda_{A,1} < \beta/\gamma, \tag{3}$$

the preventive and reactive cyber defense dynamics, namely the discrete-time version of system (1) with $\alpha = 0$, is globally exponentially stable, namely that the dynamics converges to the equilibrium $\mathbb{0}$ at an exponential speed regardless of the initial value $i(0) = [i_1(0), \ldots, i_n(0)]$.

In the more general case $\alpha \geq 0$ (i.e., preventive and reactive defenses against push-based and pull-based attacks), the following Theorem 6 supersedes the preceding Theorem 5, because condition (4) degenerates to condition (3) [34].

*Theorem 6: ([34])* Let parameters $G$, $\alpha$, $\beta$ and $\gamma$ be specified as above. Let $H = \text{diag}[h_v]_{v=1}^n$ be a diagonal matrix with

$$h_v = \left| -\beta + (1-\alpha) \prod_{u \in N_v} (1 - \gamma i_u^*) \right|.$$

If

$$\lambda_{(H+\gamma(1-\alpha)A),1} < 1, \tag{4}$$

the preventive and reactive cyber defense dynamics, namely the discrete-time version of system (1), is globally exponentially stable, namely that the dynamics converges to some equilibrium $i^* = [i_1^*, \ldots, i_n^*]$ at an exponential speed regardless of the initial value $i(0) = [i_1(0), \ldots, i_n(0)]$.

## IV. NEW RESULTS

As mentioned above, the dynamics is well understood in the parameter regime specified by the preceding conditions (3) and (4). Our goal is to characterize the dynamics when these conditions do not hold, namely when $\lambda_{A,1} \geq \beta/\gamma$ in the case $\alpha = 0$ and when $\lambda_{(H+\gamma(1-\alpha)A),1} \geq 1$ in the more general case $\alpha \geq 0$. Note that the latter degenerates to the former when $\alpha = 0$.

### A. Preparation

We first prove some basic results that are applicable to both the case $\alpha > 0$ and the case $\alpha = 0$. Denote by $f_v(i)$ the right-hand side of system (1), namely

$$f_v(i) = -\beta i_v(t) + \left[1 - (1-\alpha) \prod_{u \in N_v} (1 - \gamma i_u(t))\right](1 - i_v(t)) \tag{5}$$

where $v \in V$ and $i(t) = [i_1(t), \ldots, i_n(t)] \in [0,1]^n$ for all $v \in V$. Note that $f_v(i)$ is continuously differentiable in $[0,1]^n$. Lemma 1 below shows that system (1) with $\alpha \geq 0$ is *cooperative*.

*Lemma 1:* System (1) with $\alpha \geq 0$ is cooperative.

*Proof:* For $u \in N_v$, we have

$$\frac{\partial f_v(i)}{\partial i_u} = -(1-\alpha)(-\gamma) \prod_{w \in N_v, w \neq u} (1 - \gamma i_w(t))(1 - i_v(t))$$

$$\geq 0.$$

For $u \notin N_v$, Eq. (5) implies that $\frac{\partial f_v(i)}{\partial i_u} = 0$. It follows that

$$\frac{\partial f_v(i)}{\partial i_u} \geq 0 \quad \text{for all} \quad u \neq v \quad \text{and} \quad u, v \in V.$$

According to Definition 1, system (1) is cooperative. ∎

Lemma 2 below shows that all trajectories of system (1) are always bounded within $[0,1]^n$.

*Lemma 2:* Let $i(t)$ be the trajectory of system (1).

(a) If $\alpha > 0$, $i(t)$ will eventually enter the domain $[\epsilon, 1-\epsilon]^n$ for some constant $\epsilon \in (0, \epsilon_1^*)$ where $0 < \epsilon_1^* \leq \min\left\{\frac{1}{2}, \frac{\beta}{\beta+1}, \frac{\alpha}{\alpha+\beta}\right\}$. In other words, $i(t) \in [\epsilon, 1-\epsilon]^n$ when $t \to +\infty$.

(b) If $\alpha = 0$, $i(t)$ will eventually enter the domain $[0, 1-\epsilon]^n$ for some constant $\epsilon \in (0, \epsilon_2^*)$ where $0 < \epsilon_2^* \leq \min\left\{\frac{1}{2}, \frac{\beta}{\beta+1}\right\}$. In other words, $i(t) \in [0, 1-\epsilon]^n$ when $t \to +\infty$.

*Proof:* We prove the lemma for the case $\alpha > 0$, while noting that the case $\alpha = 0$ can be proven similarly. When $\alpha > 0$, we prove that the region $[\epsilon, 1-\epsilon]^n$ is positively invariant for some constant $\epsilon \in (0, \epsilon_1^*)$. For $v \in V$, we have

$$\left.\frac{di_v(t)}{dt}\right|_{i_v(t)=\epsilon}$$
$$= -\beta\epsilon + \left[1 - (1-\alpha) \prod_{u \in N_v} (1 - \gamma i_u(t))\right](1-\epsilon)$$
$$\geq -\beta\epsilon + \alpha(1-\epsilon)$$

and

$$\left.\frac{di_v(t)}{dt}\right|_{i_v(t)=1-\epsilon}$$
$$= -\beta(1-\epsilon) + \left[1 - (1-\alpha) \prod_{u \in N_v} (1 - \gamma i_u(t))\right]\epsilon$$
$$< -\beta(1-\epsilon) + \epsilon.$$

Then, there exists a constant $\epsilon \in (0, \epsilon_1^*)$ with $\epsilon_1^* \leq \min\left\{\frac{1}{2}, \frac{\beta}{\beta+1}, \frac{\alpha}{\alpha+\beta}\right\}$, such that

$$\left.\frac{di_v(t)}{dt}\right|_{i_v(t)=\epsilon} > 0 \quad \text{and} \quad \left.\frac{di_v(t)}{dt}\right|_{i_v(t)=1-\epsilon} < 0.$$

This implies that if $i_v(0) \in [\epsilon, 1-\epsilon]$, then $i_v(t) \in [\epsilon, 1-\epsilon]$ for all $t \in [0, +\infty)$. This means that $[\epsilon, 1-\epsilon]^n$ is positively invariant.

Observe that if $i_v(t) \leq \epsilon$, then we have

$$\frac{di_v(t)}{dt}$$
$$= -\beta i_v(t) + \left[1 - (1-\alpha) \prod_{u \in N_v} (1 - \gamma i_u(t))\right](1 - i_v(t))$$
$$\geq -\beta\epsilon + \alpha(1-\epsilon) > 0.$$

If $i_v(t) \geq 1-\epsilon$, then it holds that

$$\frac{di_v(t)}{dt}$$
$$= -\beta i_v(t) + \left[1 - (1-\alpha) \prod_{u \in N_v} (1 - \gamma i_u(t))\right](1 - i_v(t))$$
$$\leq -\beta(1-\epsilon) + \epsilon < 0.$$

Hence, $[\epsilon, 1-\epsilon]^n$ is attracting and the trajectories $i(t)$ of system (1) will eventually enter domain $[\epsilon, 1-\epsilon]^n$ regardless of the initial value. ∎

### B. Global Stability and Convergence Speed

*1) The Case $\alpha > 0$:* Lemma 2 showed that all trajectories $i(t)$ of system (1) will eventually enter an open subset of $[0, 1]^n$. This means that if system (1) has an equilibrium, the equilibrium does not belong to any boundary of $[0, 1]^n$. Theorem 7 below shows that there indeed exists a unique equilibrium in $[0, 1]^n$, and the equilibrium is neither $\mathbb{O}$ nor $\mathbb{1}$.

*Theorem 7:* Given model parameters $G$, $\alpha$, $\beta$ and $\gamma$ as specified above, let $f(i) : [0, 1]^n \to [0, 1]^n$ be the continuous map defined in Eq. (5). If $\alpha > 0$, there exists a unique equilibrium $i^* = [i_1^*, \ldots, i_n^*] \in [0, 1]^n \setminus \{\mathbb{O}, \mathbb{1}\}$ such that every trajectory of system (1) always converges to $i^*$, meaning that the preventive and reactive defense dynamics is globally stable. Moreover, the convergence is exponential for except for some zero-measure set of $(\alpha, \beta, \gamma) \in [0, 1]^3$.

**Remark**. The convergence speed proven in Theorem 7 holds except for some zero-measure set of $(\alpha, \beta, \gamma)$'s. The exclusion of this zero-measure set is necessarily for mathematical rigor, but has no practical impact because the probability that such $(\alpha, \beta, \gamma)$'s can be sampled is zero.

*2) The Case $\alpha = 0$:* As mentioned above, in the special case $\alpha = 0$, condition (4) degenerates to condition (3). Now we prove that when condition (3) does not hold, the dynamics is still globally stable.

*Theorem 8:* Given model parameters $G$, $\alpha = 0$, $\beta$ and $\gamma$ as specified above, let $f(i) : [0, 1]^n \to [0, 1]^n$ be the continuous map as defined in Eq. (5) and $\alpha = 0$.

1) In the case $\lambda_{A,1} \leq \beta/\gamma$, it holds that every trajectory $i(t)$ of system (1) converges to $\mathbb{O}$. (This supersedes the result proven in, for example, [4], [34], which applies to $\lambda_{A,1} < \beta/\gamma$ only.) When $\lambda_{A,1} < \beta/\gamma$, the convergence speed has been proven to be exponential [4], [34]; when $\lambda_1 = \beta/\gamma$, we prove here that the convergence speed is polynomial.

2) In the case $\lambda_{A,1} > \beta/\gamma$, then there exists a unique equilibrium $i^* \in [0, 1]^n \setminus \{\mathbb{O}, \mathbb{1}\}$ such that every trajectory $i(t)$ of system (1) converges to some $i^*$. Moreover, the convergence is exponential except for a zero-measure set of $(\beta, \gamma) \in [0, 1]^2$.

**Remark**. Similar to the case of Theorem 7, the exclusion of some set of zero measures is for mathematical rigor and has no practical side-effect. On the other hand, Theorem 8 also has a nice side-product. In [4], it was claimed that condition (3), namely $\lambda_{A,1} < \beta/\gamma$, is *necessary* and *sufficient* for the dynamics with $\alpha = 0$ to converge to equilibrium $\mathbb{O}$. In [34], it was shown that condition (3) is *sufficient*, but not *necessary*. The evidence given in [34] is a counter-example, but no explanation *why* condition (3) is not necessary. Theorem 8 fills the void by showing that condition (3) or $\lambda_{A,1} < \beta/\gamma$ is not necessary for the dynamics to converge to equilibrium $\mathbb{O}$, *because* the dynamics also converges to equilibrium $\mathbb{O}$ when $\lambda_{A,1} = \beta/\gamma$.

*3) Extension to the Case of Node Heterogeneity:* The global stability result established by Theorems 7-8 assume node-independent parameters, namely that $\alpha$, $\beta$ and $\gamma$ are independent of $v \in V$. While node-independent parameters may be true for uniformly defended enterprise networks, one would wonder whether or not the global stability result is still valid when the model parameters are node-dependent. This is a legitimate question because in practice, different computers may employ different sets of preventive and reactive defense mechanisms, which may have different capabilities in defending against attacks. This suggests us to consider node-dependent parameters $\alpha_v$, $\beta_v$ and $\gamma_{uv}$, which lead to the following system (6).

$$\frac{di_v(t)}{dt} = -\beta_v i_v(t) + \left[1 - (1-\alpha_v)\prod_{u \in N_v}(1-\gamma_{uv}i_u(t))\right](1-i_v(t)). \quad (6)$$

Theorem 9 below shows that system (6) is still globally stable. The proof of Theorem 9 is omitted because it is similar to the proof of Theorem 7.

*Theorem 9:* System (6) is globally stable, meaning that there exists a unique equilibrium $i^* \in [0, 1]^n$ such that every trajectory of system (6) converges to $i^*$.

### C. Bounding the Unique Equilibrium

Having proved the global stability of the dynamics in the entire parameter universe, namely the existence and uniqueness of equilibrium $i^* = [i_1^*, \cdots, i_n^*]$, it would be ideal if we can get an analytic expression of $i^*$ from

$$i_v^* = \frac{1 - (1-\alpha)\prod_{u \in N_v}(1-\gamma i_u^*)}{\beta + 1 - (1-\alpha)\prod_{u \in N_v}(1-\gamma i_u^*)}$$

for all $u, v \in V$. This turns out to be a difficult problem, except for some special cases. For example, in the special case $\alpha = 1, \beta = 1, \gamma = 1$, the system (1) can be simplified as

$$\frac{di_v(t)}{dt} = -2i_v(t) + 1, \quad v \in V.$$

The solution to this system of differential equations is:

$$i_v(t) = -\frac{1}{2}e^{-2t} + \frac{1}{2}.$$

The unique equilibrium is $i^* = [\frac{1}{2}, \cdots, \frac{1}{2}]_n$.

As an alternative, we now aim to bound the equilibrium because the bounds can be useful (e.g., the upper bound can be used in cyber defense decision-making for accommodating the worst-case scenario). To simplify the presentation, let

$$i_{\min} = \min_{v \in V} \inf_{t \in [0, +\infty)} \{i_v(t)\} \quad (7)$$

and

$$i_{\max} = \max_{v \in V} \sup_{t \in [0, +\infty)} \{i_v(t)\} \quad (8)$$

Now, we have the following theorem.

*Theorem 10:* Let $i(t) = [i_1(t), \ldots, i_n(t)]$ be the solution to system (1), $\underline{i}(t) = [\underline{i_1}(t), \cdots, \underline{i_n}(t)]$ denote the lower bound of $i(t)$ and $\overline{i}(t) = [\overline{i_1}(t), \cdots, \overline{i_n}(t)]$ denote the upper bound of $i(t)$. Then we have

$$\underline{i_v}(t) \le i_v(t) \le \overline{i_v}(t)$$

with

$$\underline{i_v}(t) = \left[i_v(0) - \frac{\mathcal{Q}_v}{\beta + \mathcal{P}_v}\right]e^{-(\beta+\mathcal{P}_v)t} + \frac{\mathcal{Q}_v}{\beta + \mathcal{P}_v} \quad (9)$$

and

$$\overline{i_v}(t) = \left[i_v(0) - \frac{\mathcal{P}_v}{\beta + \mathcal{Q}_v}\right]e^{-(\beta+\mathcal{Q}_v)t} + \frac{\mathcal{P}_v}{\beta + \mathcal{Q}_v} \quad (10)$$

where

$$\mathcal{P}_v = \begin{cases} 0, & \text{if } \alpha = 0 \text{ and } \lambda_{A,1} \le \frac{\beta}{\gamma} \\ 1 - (1-\alpha)(1-\gamma i_{\max})^{d_v}, & \\ & \text{if } (\alpha = 0 \text{ and } \lambda_{A,1} > \frac{\beta}{\gamma}) \text{ or } (\alpha > 0) \end{cases}$$

and

$$\mathcal{Q}_v = \begin{cases} 0, & \text{if } \alpha = 0 \text{ and } \lambda_{A,1} \le \frac{\beta}{\gamma} \\ 1 - (1-\alpha)\left(\frac{1}{d_v}\right)^{\frac{1}{d_v}}(1-\gamma i_{\min})^{d_v}, & \\ & \text{if } (\alpha = 0 \text{ and } \lambda_{A,1} > \frac{\beta}{\gamma}) \text{ or } (\alpha > 0) \end{cases}$$

for $v \in V$ and $t \in [0, +\infty)$.

The numerical simulation reported in Section V shows that these bounds are tighter than the bounds presented in [34]. The tightness comes from (i) knowing the global stability of the dynamics allows us to identify $i_{\min}$ and $i_{\max}$, which had to be respectively set to 0 and 1 in [34] because of the lack of information on the global stability, and (ii) we use the geometric average to make the bounds accommodate nodes' (in-)degree $\deg(v)$.

## V. SIMULATION-BASED VALIDATION

It would be ideal if we can use real data to validate the theoretical results. Unfortunately, we do not have access to such data, which is hard to obtain because of legal and privacy concerns. Instead, we use simulation to validate the results, namely the global stability of the dynamics, the relative tightness of the new upper and lower bounds of the equilibrium, and the convergence speed.

**Methodology**. In our simulation, we track the state of a node as

$$\chi_v(t) = \begin{cases} 0 & v \text{ is } \textit{secure} \text{ at time } t \\ 1 & v \text{ is } \textit{compromised} \text{ at time } t. \end{cases}$$

The rate at which $v$'s state changes from *secure* to *compromised* at time $t$ is denoted by $\tilde{\theta}_{v,0\to1}(t)$, which is a random variable dependent upon the success of pull-based and push-based attacks (because the number of *compromised* neighbors is a random variable). The rate at which $v$'s state changes from *compromised* to *secure* at time $t$ is denoted by $\tilde{\theta}_{v,1\to0}(t)$, which is dependent upon the reactive defense power. The state transition of $v$ is a Markov process with the following transition probabilities:

$$\mathbb{P}(\chi_v(t+\Delta t) = 0 | \chi_v(t))$$
$$= \begin{cases} 1 - \Delta t \cdot \tilde{\theta}_{v,0\to1}(t) + o(\Delta t), & \chi_v(t) = 0 \\ \Delta t \cdot \tilde{\theta}_{v,1\to0}(t) + o(\Delta t), & \chi_v(t) = 1 \end{cases} \quad (11)$$

and

$$\mathbb{P}(\chi_v(t+\Delta t) = 1 | \chi_v(t))$$
$$= \begin{cases} \Delta t \cdot \tilde{\theta}_{v,0\to1}(t) + o(\Delta t), & \chi_v(t) = 0 \\ 1 - \Delta t \cdot \tilde{\theta}_{v,1\to0}(t) + o(\Delta t), & \chi_v(t) = 1 \end{cases} \quad (12)$$

as $\Delta t \to 0$. It is worthy mentioning that, to the best of our knowledge, this Markov process is not tractable because of the exponential state space and the *random* (rather than *fixed*) state transition rate $\tilde{\theta}_{v,0\to1}(t)$. We describe the simulation result succinctly by plotting the average of $\sum_{v \in V} \chi_v(t)/n$ over 50 simulation runs, each of which corresponds to a random initial value $[\chi_1(0), \ldots, \chi_n(0)] \in \{0,1\}^n$. The global stability result says that all 50 runs converge to $\langle i_v(t) \rangle_{v \in V} = \sum_{v \in V} i_v(t)/n$, which can be numerically computed from Eq. (1).

**Simulation parameters.** The simulation focuses on our new results, namely the parameter regimes other than those that have been well understood according to [4], [34]. Specifically, in the case $\alpha = 0$, we focus on the parameter regimes with $\lambda_{A,1} \ge \beta/\gamma$, which violate the previously-known convergence condition (3) as discussed above; in the case $\alpha > 0$, we focus on the parameter regimes with $\lambda_{(H+\gamma(1-\alpha)A),1} \ge 1$, which violate the previously-known convergence condition (3) as discussed above. In the simulation, we set $\Delta t = 0.05$, $\tilde{\theta}_{v,1\to0}(t) = \beta$, and

$$\tilde{\theta}_{v,0\to1}(t) = 1 - (1-\alpha)\prod_{u \in N_v}(1-\gamma\chi_u(t)).$$

In each of the 50 simulation runs, we set $i_v(0)$ for $v \in V$ to be a number independently and randomly chosen from an arbitrarily chosen interval $[0.2, 0.9]$, and then determine the state $\chi_v(0)$ according to $i_v(0)$. For $G$, we use the following network structures that are available from http://snap.stanford.edu/data/. These two networks can be used as examples of cyber attack structure $G$ because attacks can indeed follow these topologies.

- The Gnutella peer-to-peer network: a directed graph with $n = 8,114$ nodes, $|E| = 26,013$ links, maximal node in-degree 61 and $\lambda_{A,1} = 4.5361$. For this network, we consider the following parameter combinations:
  (a) $\alpha = 0$, $\beta = 0.6805$ and $\gamma = 0.15$, namely, $\lambda_{A,1} = \beta/\gamma$, meaning that $\langle i_v(t) \rangle_{v \in V}$ should converge to zero.
  (b) $\alpha = 0$, $\beta = 0.8387$ and $\gamma = 0.3568$, namely, $\lambda_{A,1} > \beta/\gamma$, meaning that $\langle i_v(t) \rangle_{v \in V}$ should converge to a non-zero equilibrium.
  (c) $\alpha = 0.2456$, $\beta = 0.8159$ and $\gamma = 0.3102$, namely, $\lambda_{(H+\gamma(1-\alpha)A),1} = 1$, meaning that $\langle i_v(t) \rangle_{v \in V}$ should converge to a non-zero equilibrium.

(d) $\alpha = 0.4061$, $\beta = 0.7395$, $\gamma = 0.2012$, namely, $\lambda_{(H+\gamma(1-\alpha)A),1} > 1$, meaning that $\langle i_v(t) \rangle_{v \in V}$ should converge to a non-zero equilibrium.

- The Facebook social network: an undirected graph with $n = 4,039$ nodes, $|E| = 88,234$ links, maximal node degree $1,044$ and $\lambda_{A,1} = 162.3739$. For this network, we use the following parameter combinations:
  (a) $\alpha = 0$, $\beta = 0.4971$ and $\gamma = 0.0030$, namely, $\lambda_{A,1} = \beta/\gamma$, meaning that $\langle i_v(t) \rangle_{v \in V}$ should converge to zero.
  (b) $\alpha = 0$, $\beta = 0.8387$ and $\gamma = 0.0579$, namely, $\lambda_{A,1} > \beta/\gamma$, meaning that $\langle i_v(t) \rangle_{v \in V}$ should converge to a non-zero equilibrium.
  (c) $\alpha = 0.6535$, $\beta = 0.8656$ and $\gamma = 0.0270$, namely, $\lambda_{(H+\gamma(1-\alpha)A),1} = 1$, meaning that $\langle i_v(t) \rangle_{v \in V}$ should converge to a non-zero equilibrium.
  (d) $\alpha = 0.4361$, $\beta = 0.7395$ and $\gamma = 0.0202$, namely, $\lambda_{(H+\gamma(1-\alpha)A),1} > 1$, meaning that $\langle i_v(t) \rangle_{v \in V}$ should converge to a non-zero equilibrium.

**Confirmation of global stability and relative tightness of the bounds.** For the Gnutella network structure, Figure 5 plots the simulation result, the numerical model prediction according to system (1), the new upper bound $\overline{i}(t) = \sum_{v \in V} \overline{i_v}(t)/n$ according to Eq. (10), the new lower bound $\underline{i}(t) = \sum_{v \in V} \underline{i_v}(t)/n$ according to Eq. (9), and the upper bound $\eta$ as well as lower bound $\zeta$ given in [34]. We observe that the model prediction matches the simulation result well, while noting that the simulation result oscillates slightly. In order to see that the dynamics in each of the 50 simulation runs converge to the same equilibrium, we calculate, for each $t \in [50, 500]$, the difference $\sum_{v \in V} |\chi_v(t) - i_v(t)|/n$, where $\chi_v(t)$ is obtained from the simulation and $i_v(t)$ is numerically obtained from system (1). In principle, the difference should be close to zero in each of the simulation runs and for any $t \in [50, 500]$. In order to succinctly represent and confirm this, we first calculate for each $t \in [50, 5000]$ the standard deviations of the 50 differences $\sum_{v \in V} |\chi_v(t)/n - i_v(t)|/n$ corresponding to the 50 simulation runs, and then calculate the mean (denoted by m) and standard deviation (denoted by sd) of these standard deviations corresponding to time interval $t \in [50, 500]$. In principle, it should hold that m $\approx 0$ and sd $\approx 0$. For the parameter combinations (a)-(d) mentioned above, we respectively have m = 0.0009, 0.0139, 0.0573, 0.0319 and sd = 0.0003, 0.0027, 0.0034, 0.0021. This confirms the global stability of the dynamics. On the other hand, the new bounds are at least as tight as the previous bounds given in [34], and the new bounds are substantially tighter in most cases except for the upper bound for parameter combination (b).

For the Facebook network structure, Figure 4 plots the simulation result, the numerical model prediction according to system (1), the new upper bound $\overline{i}(t)$, the new lower bound $\underline{i}(t)$, and the upper bound $\eta$ and lower bound $\zeta$ given in [34]. We observe that the model prediction matches the simulation result well. For the parameter combinations (a)-(d) mentioned above, we report that m = 0.0011, 0.0281, 0.0424, 0.0286, and that sd = 0.0005, 0.0017, 0.0029, 0.0032, respectively.

This confirms that all of the 50 simulated dynamics converge to the same equilibrium. We also observe that the new bounds are at least at tight as the previous ones, and are substantially tighter than the previous bounds in most cases except for the upper bound in the case of parameter combination (b).

By comparing Figure 5 and Figure 4, we observe that even the new bounds for $\alpha = 0$ with $\lambda_{A,1} > \beta/\gamma$ can be very loose. This offers a great opportunity for future research: How can we characterize the circumstances under which the bounds are tight enough?

**Confirmation of convergence speed.** Theorems 7 and 8 showed that the convergence is polynomial in the case $\alpha = 0$ with $\lambda_{A,1} = \beta/\gamma$, and is exponential in all other cases. In order to confirm these results, we define the following indicator of convergence speed:

$$S(t) = \frac{1}{t} \log \left\| \frac{i(t + \Delta t) - i(t)}{\Delta t} \right\| \quad (13)$$
$$= \frac{1}{t} \log \left( \sum_{v \in V} \left| \frac{i_v(t + \Delta t) - i_v(t)}{\Delta t} \right| \right).$$

We choose this definition because $\lim_{t \to +\infty} S(t) = 0$ means the convergence is polynomial and $\lim_{t \to +\infty} S(t) = s$ for some negative constant $s$ means the convergence is exponential.

For the Gnutella network structure, Figure 6(a) plots the convergence speed given by Eq. (13) in the parameter combinations mentioned above. We observe that the convergence in case $\alpha = 0$ with $\lambda_{A,1} = \beta/\gamma$ steadily goes to 0, which confirms the polynomial convergence. For the other cases, the simulation result confirms that the convergence is exponential because Eq. (13) goes to a negative constant in these cases. Figure 6(b) plots the convergence with respect to the Facebook network structure, which exhibits similar phenomena.

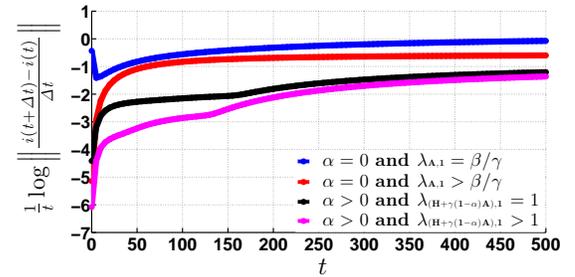

(a) Gnutella network: directed

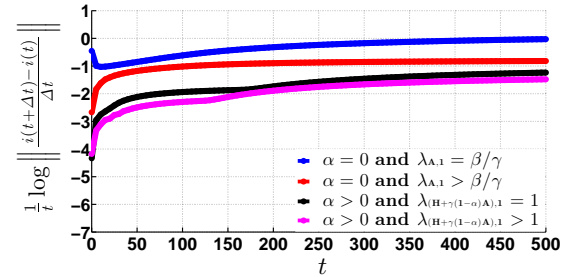

(b) Facebook network: undirected

Fig. 6. Convergence speed: polynomial vs. exponential. The smaller the value $\frac{1}{t} \log \left\| \frac{i(t+\Delta t) - i(t)}{\Delta t} \right\|$, the faster the convergence.

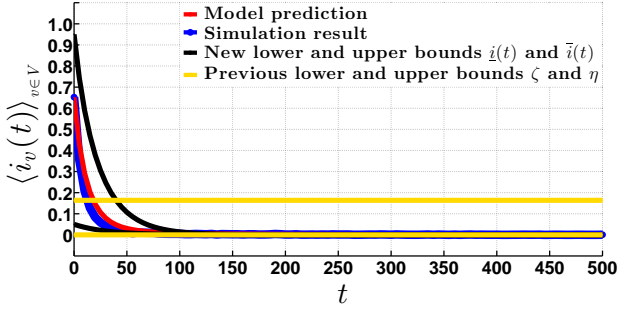
(a) $\alpha = 0$, $\beta = 0.4971$, $\gamma = 0.0030$, and $\lambda_{A,1} = \beta/\gamma$

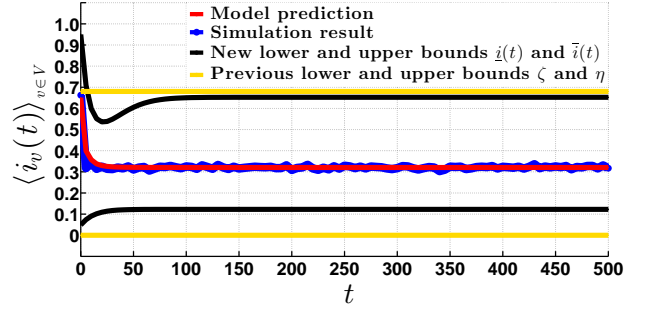
(b) $\alpha = 0$, $\beta = 0.8387$, $\gamma = 0.0579$, and $\lambda_{A,1} > \beta/\gamma$

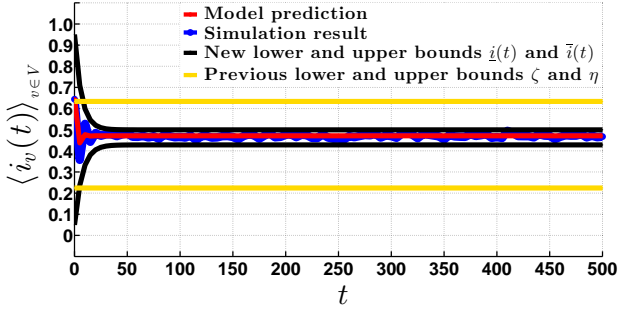
(c) $\alpha = 0.6535$, $\beta = 0.8656$, $\gamma = 0.0270$, and $\lambda_{(H+\gamma(1-\alpha)A),1} = 1$

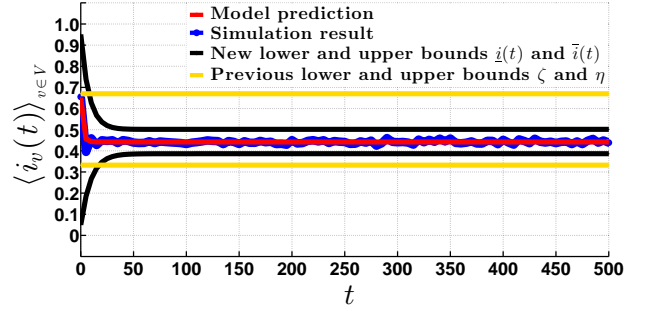
(d) $\alpha = 0.4361$, $\beta = 0.7395$, $\gamma = 0.0202$, and $\lambda_{(H+\gamma(1-\alpha)A),1} > 1$

Fig. 4. The case $G$ is the Facebook network: comparison between the simulation result, model prediction, new upper bound $\bar{i}(t)$, new lower bound $\underline{i}(t)$, and previous upper bound $\eta$ as well as lower bound $\zeta$.

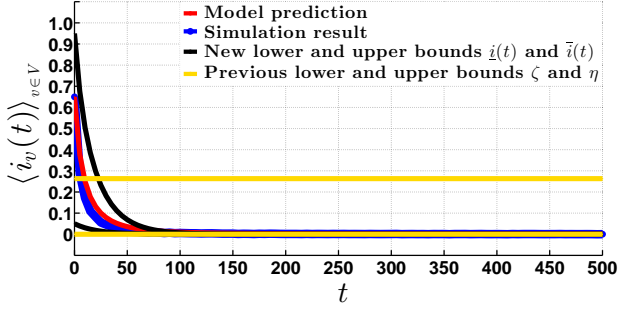
(a) $\alpha = 0$, $\beta = 0.6805$, $\gamma = 0.1500$, and $\lambda_{A,1} = \beta/\gamma$

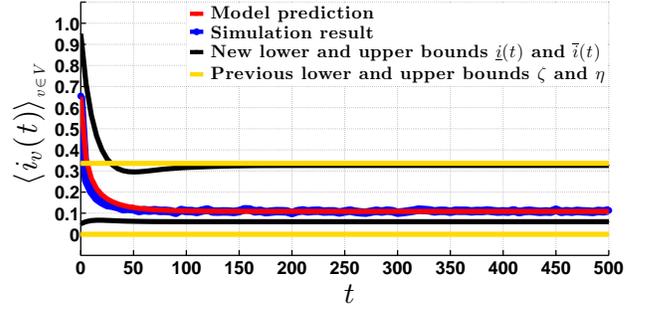
(b) $\alpha = 0$, $\beta = 0.8387$, $\gamma = 0.3568$, and $\lambda_{A,1} > \beta/\gamma$

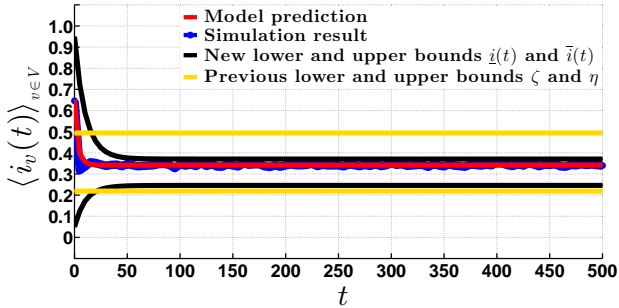
(c) $\alpha = 0.2456$, $\beta = 0.8159$, $\gamma = 0.3102$, and $\lambda_{(H+\gamma(1-\alpha)A),1} = 1$

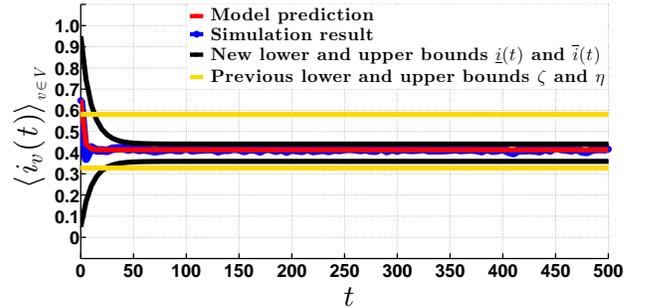
(d) $\alpha = 0.4061$, $\beta = 0.7395$, $\gamma = 0.2012$, and $\lambda_{(H+\gamma(1-\alpha)A),1} > 1$

Fig. 5. The case $G$ is the Gnutella network: comparison between the simulation result, model prediction, new upper bound $\bar{i}(t)$, new lower bound $\underline{i}(t)$, and previous upper bound $\eta$ as well as lower bound $\zeta$.

## VI. Related Work

The most closely related prior work is [34], which gave the first theoretical characterization of preventive and reactive cyber defenses against push-based and pull-based cyber attacks. The model of preventive and reactive defense dynamics was introduced in [16], which however did not give a satisfactory analytic treatment of the dynamics. The characterization presented in [34] supersedes the model of preventive and reactive cyber defenses against push-based cyber attacks (i.e., pull-based attacks were not accommodated), which has been investigated in many prior studies (e.g., [27], [7], [4]). In particular, [27] is the first paper that established the connection between the fate of the dynamics in *arbitrary* cyber attack structures and their algebraic properties. Our results supersede the ones presented in [34], and therefore the results presented in [27], [7], [4]. In particular, we show that the dynamics of preventive and reactive defenses against push- and pull-based attacks is globally stable in the entire parameter universe that has been investigated in the literature, which gives a full characterization of the dynamics.

In a broader context, preventive and reactive defense dynamics is a specific kind of cybersecurity dynamics [30], [31]. In parallel to preventive and reactive defense dynamics, the following kinds of dynamics have been studied recently: adaptive defense dynamics [35], active defense dynamics [39], [33], [18], and proactive defense dynamics [8]; Deeply understanding these kinds of dynamics will pave the way for establishing a unified framework that accommodates all kinds of attacks and all kinds of defenses [30].

## VII. Conclusion

We have shown that preventive and reactive defense dynamics is globally stable in the entire parameter universe, and discussed its cybersecurity meanings. Our characterization of the convergence speed has led to a complete understanding on this matter in the entire parameter universe. We have presented new bounds of the equilibrium, and showed that these bounds are tighter than the ones given in the literature.

We have discussed several exciting directions for future research in Section II. In addition, we anticipate that the dynamics is also globally stable in the discrete-time model [34]. Due to space limitation, we will give a detailed analysis elsewhere.

## APPENDIX

*Proof:* (proof of Theorem 7) Consider the continuous map $g(i) : [0,1]^n \to [1, 1+\rho]^n$ with $g(i) = 1 + \rho f(i)$ and $\rho > 0$ being a constant. For any $\delta \in (0,1)$, $i \in [0,1]^n \backslash \{\mathbb{O}\}$ and $\alpha > 0$, we have

$$g_v(\delta i) - \delta g_v(i) = (1-\delta) + \rho f(\delta i) - \delta \rho f(i)$$
$$= (1-\delta) + \rho \bigg\{ -\beta \delta i_v(t) +$$
$$\bigg[1 - (1-\alpha) \prod_{u \in N_v}\Big(1-\gamma\delta i_u(t)\Big)\bigg]\big(1-\delta i_v(t)\big) + \beta\delta i_v(t)$$
$$-\delta\bigg[1 - (1-\alpha) \prod_{u \in N_v}\Big(1-\gamma i_u(t)\Big)\bigg]\big(1-i_v(t)\big)\bigg\}$$
$$> (1-\delta) + \rho\big(1-i_v(t)\big)\bigg\{\bigg[1-(1-\alpha)\prod_{u\in N_v}\Big(1-\gamma\delta i_u(t)\Big)\bigg]$$
$$-\delta\bigg[1 - (1-\alpha) \prod_{u \in N_v}\Big(1-\gamma i_u(t)\Big)\bigg]\bigg\}$$
$$= (1-\delta) + \rho\big(1-i_v(t)\big)\bigg\{(1-\delta) - (1-\alpha) \times$$
$$\bigg[\prod_{u \in N_v}\Big(1-\gamma\delta i_u(t)\Big) - \delta\prod_{u \in N_v}\Big(1-\gamma i_u(t)\Big)\bigg]\bigg\}$$
$$> 0 \quad \text{/* by using } \delta \in (0,1) \text{ and inequality (14) below */}$$

for all $v \in V$. This means $g(\delta i) > \delta g(i)$ for any $\delta \in (0,1)$. According to Definition 2, $g(i)$ is strongly subhomogeneous on $[0,1]^n$. This means condition (b) required by Theorem 2 is satisfied. Since system (1) is cooperative, Theorem 1 says $f(i)$ is monotone on $[0,1]^n$, which means $g(i)$ is monotone on $[0,1]^n$. This means condition (a) required by Theorem 2 is satisfied. Lemma 2 says the domain $[\epsilon, 1-\epsilon]^n$ is a nonempty compact invariant subset of $[0,1]^n$, and Theorem 2 says that $[\epsilon, 1-\epsilon]^n$ contains a single equilibrium. Because $g(i) = 1 + \rho f(i)$ and $f(i)$ is a continuously monotone function on $[\epsilon, 1-\epsilon]^n$, $f(i)$ admits only one equilibrium $i^* \gg \mathbb{O}$ in $[\epsilon, 1-\epsilon]^n$. Theorem 3 says that every trajectory attracted to the closed domain $[\epsilon, 1-\epsilon]^n$ that contains only one equilibrium $i^*$ converges to $i^*$. Lemma 2 says $i(t)$ will eventually enter the domain $[\epsilon, 1-\epsilon]^n$. Hence, we conclude that there exists a unique equilibrium $i^* \in [0,1]^n \backslash \{\mathbb{O}, \mathbb{1}\}$ such that every trajectory of system (1) always converges to $i^*$.

The remaining task is to prove the following inequality (14):

$$(1-\delta) - (1-\alpha) \times$$
$$\bigg[\prod_{u \in N_v}\Big(1-\gamma\delta i_u(t)\Big) - \delta\prod_{u \in N_v}\Big(1-\gamma i_u(t)\Big)\bigg] > 0. \quad (14)$$

In what follows we prove inequality (14) for any node $v$, by induction on $\deg(v)$.

When $\deg(v) = 1$, we have

$$(1-\delta) - (1-\alpha)\bigg[\Big(1-\gamma\delta i_u(t)\Big) - \delta\Big(1-\gamma i_u(t)\Big)\bigg]$$
$$= \big(1 - i_v(t)\big)(1-\delta)\alpha > 0.$$

Suppose inequality (14) holds when $\deg(v) = d_v$ with $1 \leq d_v < n$, namely,

$$(1-\delta) -$$
$$(1-\alpha)\bigg[\prod_{u \in N_v}\Big(1-\gamma\delta i_u(t)\Big) - \delta\prod_{u \in N_v}\Big(1-\gamma i_u(t)\Big)\bigg] > 0.$$

When $\deg(v) = d_v + 1$, we have

$$(1-\delta) - (1-\alpha) \times$$
$$\bigg[\prod_{u \in N_v \cup \{w\}}\Big(1-\gamma\delta i_u(t)\Big) - \delta\prod_{u \in N_v \cup \{w\}}\Big(1-\gamma i_u(t)\Big)\bigg]$$
$$= (1-\delta) - (1-\alpha) \times$$
$$\bigg[\prod_{u \in N_v}\Big(1-\gamma\delta i_u(t)\Big) - \delta\prod_{u \in N_v}\Big(1-\gamma i_u(t)\Big)\bigg] + (1-\alpha) \times$$
$$\bigg[\prod_{u \in N_v}\Big(1-\gamma\delta i_u(t)\Big)\gamma\delta i_w(t) - \delta\prod_{u \in N_v}\Big(1-\gamma i_u(t)\Big)\gamma i_w(t)\bigg]$$
$$> (1-\alpha)\bigg[\prod_{u \in N_v}\Big(1-\gamma\delta i_u(t)\Big) - \prod_{u \in N_v}\Big(1-\gamma i_u(t)\Big)\bigg]\gamma\delta i_w(t)$$
$$> 0$$

where $w \neq u$ and $u, w \in N_v$. This means inequality (14) holds when $\deg(v) = d_v + 1$. In other words, inequality (14) holds for every $\deg(v)$.

Now we prove that the convergence is exponential. We observe that the largest real part of all eigenvalues of the Jacobin matrix $[\frac{\partial f_v}{\partial i_u}]_{v,u=1}^n$ at equilibrium $i = i^*$ should be non-positive (otherwise, $i^*$ is unstable). Let

$$G(\alpha) = \max\bigg\{\mathcal{R}e(\lambda) : \lambda \in \sigma\bigg(\Big[\frac{\partial f_v}{\partial i_u}\Big]_{v,u=1}^n\bigg)\bigg\},$$

where $\sigma(\cdot)$ stands for the set of eigenvalues of the Jacobin matrix and $\mathcal{R}e(\cdot)$ stands for the real parts of the eigenvalues. Lemma 1 showed that system (1) is cooperative, meaning that all off-diagonal elements of the Jacobin matrix are nonnegative and $G(\alpha)$ is one of the eigenvalues. Since $G(\alpha)$ is an analytic function with respect to $\alpha$, the solutions of $G(\alpha) = 0$ is finite in $(0,1]$. This means that the set $W_\alpha$ of solutions of $G(\alpha) = 0$ has a zero measure. If $\alpha \notin W_\alpha$, then $G(\alpha) < 0$, meaning that the convergence towards $i^*$ is exponential. The same reasoning applies to $\beta$ and $W_\beta$ as well as $\gamma$ and $W_\gamma$. Putting these together, we conclude that the dynamics converges to $i^*$ exponentially except for a zero-measure set of $(\alpha, \beta, \gamma) \in W_\alpha \times W_\beta \times W_\gamma \subset [0,1]^3$. This completes the proof. ∎

*Proof:* (proof of Theorem 8) When $\alpha = 0$, it is clear that $i^* = \mathbb{O}$ is an equilibrium of system (1).

If $\lambda_{A,1} \leq \beta/\gamma$, Lemma 1 says that system (1) with $\alpha = 0$ is cooperative. Lemma 2 tells that every trajectory $i(t)$ of system (1) with $\alpha = 0$ will eventually enter the bounded domain $[0, 1-\epsilon]^n$ for some constant $\epsilon \in (0, \epsilon_2^*)$ where $0 < \epsilon_2^* \leq \frac{1}{2}$. The first result in Theorem 4 as well as its proof confirms that every trajectory $i(t)$ of system (1) will converge to equilibrium $\mathbb{O}$.

Now we prove the convergence is polynomial when $\lambda_{A,1} = \beta/\gamma$. We observe that all off-diagonal elements of the Jacobin matrix at equilibrium $i = i^*$ are nonnegative. Without loss of generality, we suppose that the geometrical dimension of eigenvalue $\lambda_{A,1}$ is 1, because it is straightforward to deal with the case that the geometrical dimension is greater than 1. This means that the right eigenvector associated with $\lambda_{A,1}$ has all components nonnegative, which can be denoted by $\xi = [\xi_1, \cdots, \xi_n]^\top$ with $\sum_{v=1}^n \xi_v = 1$. Let $j(t) = i(t) - i^*$ be the variation at $i^*$, which exponentially converges to zero in the direction of $\xi$. Consider $j(t) = \eta(t)\xi$ for some scalar function $\eta(t)$. From the Taylor's expansion of system (1) at equilibrium $i = i^*$ up to the second order, we have, for $v \in V$,

$$\frac{d\eta(t)}{dt}\xi_v = \sum_{u=1}^n \frac{\partial f_v}{\partial i_u}(i^*)\xi_u \eta(t) - c_v^* \eta^2(t) + O(\eta^3(t))$$
$$= -c_v^* \eta^2(t) + O(\eta^3(t)) \quad (15)$$

with

$$c_v^* = -2(1-\alpha)\gamma \sum_{w \in N_v} \prod_{\substack{u' \neq w \\ u' \in N_v}} (1 - \gamma i_{u'}^*)\xi_w \xi_v$$
$$- (1 - i_v^*)(1-\alpha)\gamma^2 \sum_{\substack{u, w \neq v \\ u, w \in N_v}} \prod_{\substack{u' \notin \{u, w\} \\ u' \in N_v}} (1 - \gamma i_{u'}^*)\xi_u \xi_w.$$

Summing equations (15) over $v$, we have

$$\frac{d\eta(t)}{dt} = -c^*\eta^2(t) + O(\eta^3(t))$$

with $c^* = \sum_{v=1}^n c_v^*$ is a positive constant. This means that the convergence of $\eta(t)$ towards zero is polynomial near equilibrium $i^*$. This completes the proof of the first statement in Theorem 8.

If $\lambda_{A,1} > \beta/\gamma$, we first consider the case that $G$ is strongly connected and then extend the analysis to the case that $G$ has an arbitrary topology.

In the case $G$ is strongly connected, we observe that condition (a) of Theorem 4 is satisfied. Similar to the algebra used in the proof in Theorem 7, we can use induction to prove

$$f_v(\delta i) - \delta f_v(i)$$
$$= -\beta \delta i_v + (1 - \prod_{u \in N_v}(1 - \delta\gamma i_u))(1 - \delta i_v)$$
$$- \delta\left[-\beta i_v + (1 - \prod_{u \in N_v}(1 - \gamma i_u))(1 - i_v)\right] \quad (16)$$
$$> (1 - i_v(t))\left\{(1-\delta) - \left[\prod_{u \in N_v}\left(1 - \gamma\delta i_u(t)\right)\right.\right.$$
$$\left.\left. - \delta\prod_{u \in N_v}\left(1 - \gamma i_u(t)\right)\right]\right\}$$
$$> 0$$

when there exists $v \in V$ such that $\deg(v) \geq 2$. When $\deg(v) \leq 1$ for all $v \in V$, it can be verified directly from Eq. (16) that $f_v(\delta i) > \delta f_v(i)$ still holds. Note that $[\partial f_v(i)/\partial i_u]_{u,v \in V_{p-1}}$ is irreducible and $f_v(i)$ is strictly sub-homogeneous for $v \in V_{p-1}$ and $i \gg \mathbb{O}$. By Theorem 4, we have $\lim_{t \to \infty} i(t) = i^*$ for all initial values in $(0, 1]^n$, where $i^* \gg \mathbb{O}$.

In the case $G$ is not strongly connected, $G$ can be partitioned into some strongly connected components according to the Frobenius form (2). This means that there are two cases: $p = 2$ and $p > 2$.

When $p = 2$, $A$ has the following form:

$$A = \begin{bmatrix} A_{11} & A_{12} \\ 0 & A_{22} \end{bmatrix}$$

where $A_{12}$ is not a zero matrix. For an arbitrary strongly connected component $V_2$ without in-edges, if $V_2$ is singlet, then it is trivial to prove that $i_{V_2}$ converges to some positive value for all $v \in V_j$; otherwise, following the same algebras above, under the conditions that $deg(v) \geq 2$ for some $v \in V_2$, one can prove that $i_v(t)$ converges to some positive value, $i_v^* > 0$, for all $v \in V_2$. Then, we can rewrite Eq. (1) restricted on $V_1$ as follows

$$\frac{di_v(t)}{dt} = -\beta i_v(t) + \left[1 - \prod_{u \in N_v \cap V_1}(1 - \gamma i_u(t))\right.$$
$$\left. \cdot \prod_{u \in N_v \cap V_2}(1 - \gamma i_u(t))\right](1 - i_v(t)) \quad (17)$$

for $v \in V_1$. By assigning $i_v(t)$ as $i_v^*$ for each $v \in V_2$, Eq. (17) asymptotically becomes

$$\frac{di_v(t)}{dt} = -\beta i_v(t) + \left[1 - c^*\prod_{u \in N_v \cap V_1}(1 - \gamma i_u(t))\right](1 - i_v(t))$$
$$(18)$$

for $v \in V_1$ with $c^* = \prod_{u \in N_v \cap V_2}(1 - \gamma i_u^*)$ being a positive constant less than 1. If $\lambda_{A_{11},1} \leq \beta/\gamma$, we already proved in the first case of the present theorem that subsystem (18) converges. If $\lambda_{A_{11},1} > \beta/\gamma$, since $A_{11}$ is a strongly connected component, we can also obtain that subsystem (18) converges by using the same method as for proving that $i_v(t)$ converges to $i_v^*$ for all $v \in V_2$.

When $p > 2$, we can use induction on $p$ to prove that $i_v(t)$ converges to $i_v^*$ for all $v \in V$.

The exponential convergence of the dynamics under the condition $\lambda_{A,1} > \beta/\gamma$ can be proven similarly to the proof of the convergence speed in Theorem 7. In other words, except some zero-measure set of $(\beta, \gamma) \in [0, 1]^2$, the convergence for system (1) with $\lambda_{A,1} > \beta/\gamma$ is exponential. This completes the proof of the second statement in Theorem 8. ∎

*Proof:* (proof of Theorem 10) For each node $v \in V$,

consider the following derivatives:

$$\frac{di_v(t)}{dt}$$
$$= -\beta i_v(t) + \left[1 - (1-\alpha)\prod_{u \in N_v}(1 - \gamma i_u(t))\right](1 - i_v(t))$$
$$\geq -\beta i_v(t) + \left\{1 - (1-\alpha)\left[\frac{1}{d_v}\sum_{u \in N_v}(1 - \gamma i_u(t))\right]^{d_v}\right\}(1 - i_v(t))$$
$$\geq -\beta i_v(t) - \left[1 - (1-\alpha)(1 - \gamma i_{\max})^{d_v}\right]i_v(t) + \left[1 - (1-\alpha)\left(\frac{1}{d_v}\right)^{d_v}\sum_{u \in N_v}(1 - \gamma i_u(t))^{d_v}\right]$$
$$\geq -\left[\beta + 1 - (1-\alpha)(1 - \gamma i_{\max})^{d_v}\right]i_v(t) + \left[1 - (1-\alpha)\left(\frac{1}{d_v}\right)^{d_v - 1}(1 - \gamma i_{\min})^{d_v}\right]$$
$$\geq -\left[\beta + 1 - (1-\alpha)(1 - \gamma i_{\max})^{d_v}\right]i_v(t) + \left[1 - (1-\alpha)\left(\frac{1}{d_v}\right)^{\frac{1}{d_v}}(1 - \gamma i_{\min})^{d_v}\right],$$

and

$$\frac{di_v(t)}{dt}$$
$$= -\beta i_v(t) + \left[1 - (1-\alpha)\prod_{u \in N_v}(1 - \gamma i_u(t))\right](1 - i_v(t))$$
$$\leq -\beta i_v(t) - \left\{1 - (1-\alpha)\left[\frac{1}{d_v}\sum_{u \in N_v}(1 - \gamma i_u(t))\right]^{d_v}\right\}i_v(t) + \left[1 - (1-\alpha)(1 - \gamma i_{\max})^{d_v}\right]$$
$$\leq -\beta i_v(t) - \left[1 - (1-\alpha)\left(\frac{1}{d_v}\right)^{d_v}\sum_{u \in N_v}(1 - \gamma i_u(t))^{d_v}\right]i_v(t) + \left[1 - (1-\alpha)(1 - \gamma i_{\max})^{d_v}\right]$$
$$\leq -\left[\beta + 1 - (1-\alpha)\left(\frac{1}{d_v}\right)^{d_v-1}(1 - \gamma i_{\min})^{d_v}\right]i_v(t) + \left[1 - (1-\alpha)(1 - \gamma i_{\max})^{d_v}\right]$$
$$\leq -\left[\beta + 1 - (1-\alpha)\left(\frac{1}{d_v}\right)^{\frac{1}{d_v}}(1 - \gamma i_{\min})^{d_v}\right]i_v(t) + \left[1 - (1-\alpha)(1 - \gamma i_{\max})^{d_v}\right].$$

For any $v \in V$, we observe the following:

$$\begin{cases} \dfrac{di_v(t)}{dt} \geq -(\beta - \mathcal{P}_v)i_v(t) + \mathcal{Q}_v \\ \dfrac{di_v(t)}{dt} \leq -(\beta - \mathcal{Q}_v)i_v(t) + \mathcal{P}_v \end{cases}$$

where

$$\mathcal{P}_v = \begin{cases} 0, & \text{if } \alpha = 0 \text{ and } \lambda_{A,1} \leq \frac{\beta}{\gamma} \\ 1 - (1-\alpha)(1 - \gamma i_{\max})^{d_v}, & \\ & \text{if } (\alpha = 0 \text{ and } \lambda_{A,1} > \frac{\beta}{\gamma}) \text{ or } (\alpha > 0) \end{cases},$$

$$\mathcal{Q}_v = \begin{cases} 0, & \text{if } \alpha = 0 \text{ and } \lambda_{A,1} \leq \frac{\beta}{\gamma} \\ 1 - (1-\alpha)\left(\dfrac{1}{d_v}\right)^{\frac{1}{d_v}}(1 - \gamma i_{\min})^{d_v}, & \\ & \text{if } (\alpha = 0 \text{ and } \lambda_{A,1} > \frac{\beta}{\gamma}) \text{ or } (\alpha > 0) \end{cases}.$$

with $i_{\min}$ and $i_{\max}$ defined in (7) and (8).

According to the Gronwall inequality [25], we obtain $\underline{i_v}(t) \leq i_v(t) \leq \overline{i_v}(t)$ where

$$\underline{i_v}(t) = \left[i_v(0) - \frac{\mathcal{Q}_v}{\beta + \mathcal{P}_v}\right]e^{-(\beta + \mathcal{P}_v)t} + \frac{\mathcal{Q}_v}{\beta + \mathcal{P}_v}$$

and

$$\overline{i_v}(t) = \left[i_v(0) - \frac{\mathcal{P}_v}{\beta + \mathcal{Q}_v}\right]e^{-(\beta + \mathcal{Q}_v)t} + \frac{\mathcal{P}_v}{\beta + \mathcal{Q}_v}$$

for any $v \in V$ and all $t \in [0, +\infty)$. This completes the proof. ∎